\documentclass[twocolumn]{aastex6} 
\usepackage{bm}
\usepackage{amsmath,amssymb}

\usepackage{graphicx}
\usepackage{color}

\shorttitle{Multiple BH merger concordant with GW events}
\shortauthors{Tagawa and Umemura}

\begin{document}

\title{Merger of multiple accreting black holes concordant with gravitational wave events}

\author{
	{Hiromichi Tagawa}\altaffilmark{1,2} and
{Masayuki Umemura}\altaffilmark{3}
}
\affil{
	\altaffilmark{1}{Institute of Physics, E{\"o}tv{\"o}s University, P{\'a}zm{\'a}ny P.s., Budapest, 1117, Hungary}; {htagawa@caesar.elte.hu}\\
\altaffilmark{2}{National Astronomical Observatory of Japan, 2-21-1 Osawa, Mitaka, Tokyo 181-8588, Japan}\\
\altaffilmark{3}{Center for Computational Sciences, University of Tsukuba, Tsukuba, Ibaraki 305-8577, Japan}
}

\begin{abstract}
Recently, advanced Laser Interferometer Gravitational-Wave Observatory (aLIGO) 
has detected black hole (BH) merger events, most of which are sourced by 
BHs more massive than $30~M_\odot$.  
Especially, the observation of GW170104 suggests dynamically assembled binaries 
favoring a distribution of misaligned spins.
It has been argued that mergers of unassociated BHs can be engendered 
through a "chance meeting" in a multiple BH system under gas-rich environments.
In this paper, we consider the merger of unassociated BHs, concordant with the massive BH merger events. 
To that end, we simulate a multiple BH system 
with a post-Newtonian $N$-body code incorporating gas accretion and general relativistic effects. 
As a result, we find that 
gas dynamical friction effectively promotes three-body interaction of BHs 
 in dense gas of $n_\mathrm{gas}\gtrsim 10^6 ~\mathrm{cm}^{-3}$,
so that BH mergers can take place within  $30$ Myr. 
This scenario predicts an isotropic distribution of spin tilts.
In the concordant models with GW150914, 
the masses of seed BHs are required to be $\gtrsim25M_\odot$.
The potential sites of such "chance meeting" BH mergers are active galactic nucleus (AGN) disks
and dense interstellar clouds. Assuming the LIGO O1,  
we roughly estimate the event rates for PopI BHs and PopIII BHs in AGN disks to be respectively
$\simeq 1-2~\mathrm{yr}^{-1}$ and $\simeq 1~\mathrm{yr}^{-1}$.
Multiple episodes of AGNs may enhance the rates by roughly 
an order of magnitude. 
For massive PopI BHs in dense interstellar clouds, the rate is $\simeq 0.02~\mathrm{yr}^{-1}$. 
Hence, high-density AGN disks are a more plausible site for
mergers of chance meeting BHs. 
\end{abstract}
\keywords{
galaxies: active 
-- gravitational waves 
-- ISM: clouds 
-- methods: numerical 
-- stars: black holes 
-- stars: Population III
}

\section{Introduction}

Recently, gravitational wave (GW) emission
associated with black hole (BH) mergers has been detected by
advanced LIGO (aLIGO), 
in the events of
GW150914 \citep{GW150914},
GW151226 \citep{GW151226},
GW170104 \citep{GW170104}, 
GW170608 \citep{GW170608}, and
GW170814 \citep{GW170814}. 
Excepting the GW151226 and GW170608 events, the BH pair in each event includes 
a BH more massive than $30~M_{\odot}$. 
\citet{Abbott16c} have argued that 
such massive BHs are unlikely to originate in metal-rich stars 
owing to mass loss by stellar wind. 
As models for the BH merger events, several binary evolution scenarios  
have been proposed. 
They include a binary of metal-free or low-metallicity stars
accompanied by mass transfer or common envelope ejection 
\citep*[e.g.][]{Kinugawa14,Belczynski16}, 
binary evolution in a tidally distorted field \citep[e.g.][]{deMink16}, 
binary evolution driven by fallback accretion \citep{Tagawa18}
and dynamical interaction in dense stellar clusters \citep*[e.g.][]{OLeary09,Samsing14,Rodriguez16}.
Also, BH binaries may be hardened within gas-rich environments 
\citep*{Escala04,Chapon13}, 
especially in active galactic nucleus (AGN) disks \citep*{Kocsis11,McKernan12,McKernan14,Bartos17,Stone17,McKernan17}. 
\citet*{McKernan12,McKernan14} predicted the occurrence of intermediate BH mergers originating in AGN disks. 
\citet{McKernan17} also considered 
binary formation of unassociated BHs through angular momentum exchange. 
\citet{Baruteau11} investigated inward migration of massive stellar binaries hardened
whitin a dense gaseous disk in the Galactic center. 

GW observations can provide information about component spins 
through measurements of an effective inspiral spin parameter $\chi_{\rm eff}$,
which can potentially be used to distinguish different formation channels. 
Isolated binary evolution does not result in a significant spin misalignment,
since mass transfer and tides are to align spins with the orbital angular momentum. 
The GW170104 event exhibits
$\chi_{\rm eff}=-0.12^{+0.21}_{-0.30}$, which disfavors
spin configurations with both component spins positively aligned with the orbital
angular momentum \citep{GW170104}, 
although the less massive BH merger in GW151226 has a preference for spins
with positive projections along the orbital angular momentum \citep*{GW151226}. 
The observation of GW170104 hints towards dynamically assembled binaries 
favoring a distribution of misaligned spins rather than near orbit-aligned spins.
Recently, \citet{tag15,tag16} have proposed mergers of unassociated BHs
through a "chance meeting" in gas-rich environments, 
without making a priori assumption of a BH binary. 
They have demonstrated that a multiple stellar-mass BH system embedded in dense gas
can engender mergers of BHs through gas dynamical friction and three-body interaction, 
which predicts an isotropic distribution of spin tilts.

In this paper, we consider BH mergers by {\it chance meetings} 
in a multiple BH system, 
especially focusing on the massive BH merger events 
(GW150914, GW170104, and GW170814).
Favorable gas-rich environments for BH mergers are provided 
in nuclear regions of galaxies which have
density of $n_\mathrm{gas}\gtrsim10^{7}~\mathrm{cm}^{-3}$ at
$\lesssim 1~\mathrm{pc}$  \citep*{Goodman03,Namekata16}. 
Another possible site is
dense interstellar cloud cores of $n_\mathrm{gas}=10^{5-7}~\mathrm{cm}^{-3}$ \citep{Bergin96,Stahler10}, 
or interstellar clouds of $n_\mathrm{gas}<10^{5}~\mathrm{cm}^{-3}$ \citep{Spitzer78}. 
Simulations are performed with a highly accurate post-Newtonian $N$-body code, 
where such general relativistic effects as the pericenter shift and GW emission 
are taken into consideration.
In these simulations, the effects of gas dynamical friction and Hoyle-Lyttleton mass 
accretion by ambient gas are incorporated. 
Changing initial masses of BHs, ambient gas density, and
distributions of BHs, we derive the range of BH mass that is concordant with
the GW events, and thereby assess the mass of accreting gas before mergers. 
Also, we roughly estimate the event rates of such BH mergers 
both in galactic centers and in dense interstellar clouds.

\section{Post-Newtonian $N$-body Simulations}

\subsection{Numerical scheme}

The detailed description of numerical schemes is given in \citet{tag16}. 
The equations of motion are integrated using a fourth-order Hermite scheme \citep{mak92}. 
Our simulations incorporate the effects of gas dynamical friction and
gas accretion onto BHs. The general relativistic effects are dealt with 
post-Newtonian prescription up to a 2.5PN term \citep{Kupi06}, where
1PN and 2PN terms correspond to pericentre shift, and a 2.5PN term to GW emission. 

\subsection{Setup of Simulations}

The key parameters in our simulations are initial BH mass ($m_0$), 
initial typical extension of BH distributions ($r_\mathrm{typ}$),
ambient gas number density ($n_\mathrm{gas}$), 
and accretion efficiency ($\epsilon$). 
We set gas accretion rate onto each BH to be 
the accretion efficiency ($\epsilon\leq1$) times 
the Hoyle-Lyttleton accretion rate  ($\dot{m}_{\mathrm{HL}}$), i.e., 
\begin{equation}
	\dot{m}_{i}=\epsilon\dot{m}_{\mathrm{HL},i}=\epsilon\frac{4\pi G^2 m_\mathrm{H} n_\mathrm{gas}m_{i}^2}{(c_\mathrm{s}^2+v_i^2)^{3/2}}, 
\label{eq_rhoyle}
\end{equation}
where $v_i$ is velocity of $i$-th BH, $c_{\rm s}$ is sound speed, 
$G$ is the Gravitational constant, and $m_{\rm H}$ is the hydrogen mass. 
The effect of radiation pressure on Hoyle-Lyttleton accretion \citep*{wat00,han01} is incorporated as Tagawa et al. (2016). 
This gas accretion prescription allows super-Eddington accretion, 
which is verified in spherical symmetric systems \citep[e.g.][]{Inayoshi16}. 
We consider multiple BHs that are embedded in high-density gas, e.g.,
in galactic nuclear regions of $\lesssim1$ pc or in dense interstellar clouds. 
Then, typical extensions of BH distributions 
at an initial epoch ($r_\mathrm{typ}$) are assumed to be from $0.01$ to $1$ pc. 
Additionally, to scrutinize dependence on ambient gas density, 
we consider a relatively wide range of gas density 
from $10^2~\mathrm{cm^{-3}}$ to $10^{10}~\mathrm{cm^{-3}}$. 
We initially set up five BHs with equal mass of 20, 25, or 30 $M_\odot$. 
Because of the uncertainty concerning the actual mass accretion rate,  
	we vary the gas accretion efficiency $\epsilon$ in a range of $10^{-3}$ to $1$. 


We set BHs in a uniform gas sphere whose mass is $10^5~{M_\odot}$. 
Therefore, according to the choice of gas density, the radius of gas sphere, $R_\mathrm{gas}$, 
is changed.
The gas temperature is assumed to be $1000~\mathrm{K}$ (therefore $c_{\rm s} = 3.709~{\rm km~s}^{-1}$) as \citet{tag16}. 
Initial positions of BHs are set randomly in a $x-y$ plane within 
$r_\mathrm{typ}$ which is smaller than $R_\mathrm{gas}$. 
Initial velocity of each BH is given as the sum of a circular component and a random component. 
Circular velocity is given so that the centrifugal force should balance the gravity by gas in the $x-y$ plane. 
In addition, we impose random velocities in the $xyz$ space, 
according to the probability of a Gaussian distribution with the same dispersion as the circular velocity. 

We adjudicate that two BHs merge, when their separation is less than 100 times 
the sum of their Schwarzschild radii. 
The evolution is pursued for $10~\mathrm{Gyr}$, 
since we consider mergers within the cosmic time. 
We terminate the simulation, when the first BH merger occurs.

\begin{figure}
\includegraphics[width=90mm]{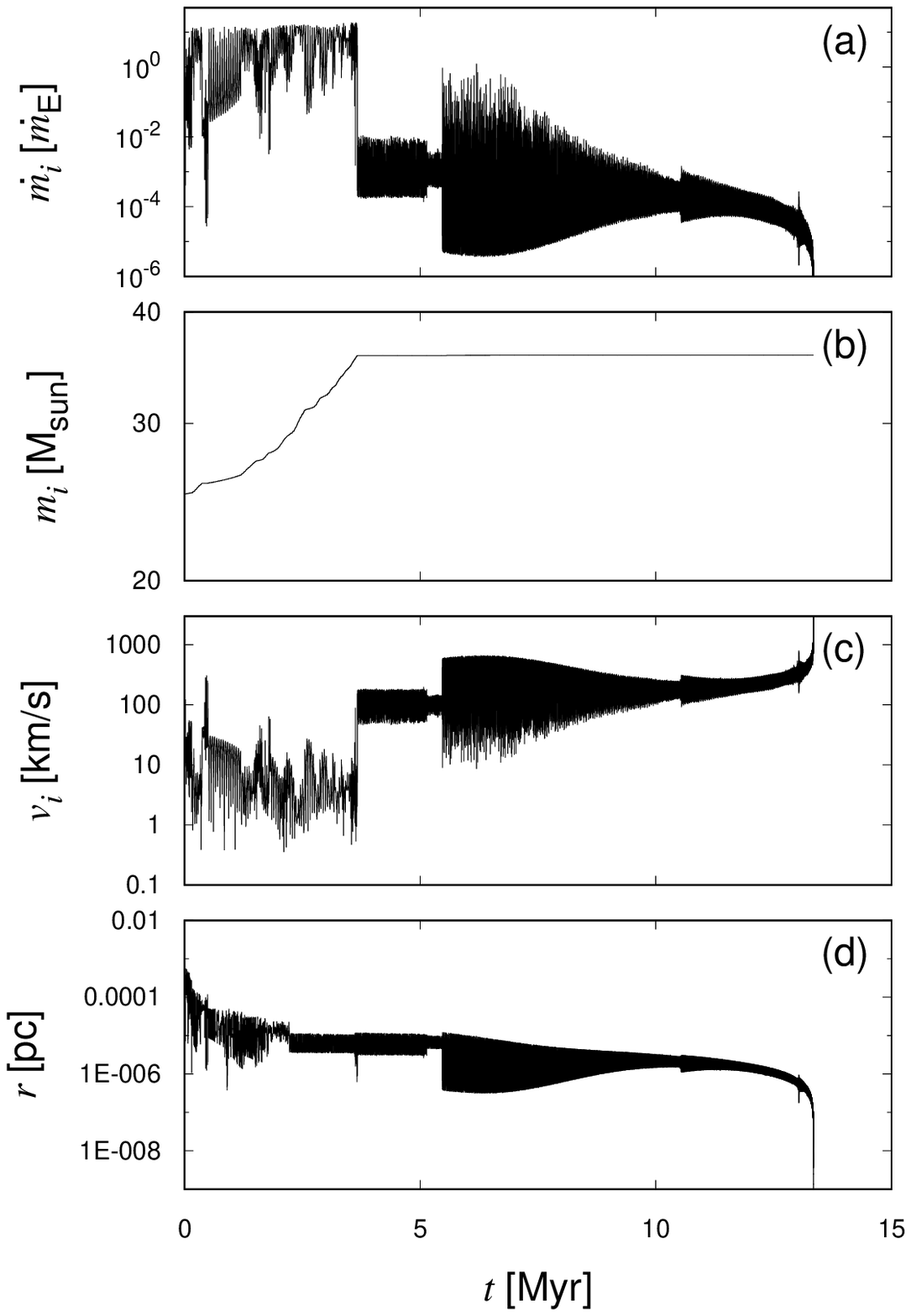}
\caption{
	The time evolution of physical quantities for Model $3$ in Table 1. 
	Panels (a), (b), and (c) represent mass accretion rate in units of
	the Eddington mass accretion rate ($\dot{m}_\mathrm{E}=L_\mathrm{E}/c\eta$,$\eta=0.1$), mass,
	and velocity for a heavier BH in merged BHs, respectively. 
	Panel (d) shows the separation of the closest pair within all BHs. 
}
	\label{vtall23}
\end{figure}

\begin{table*}
	\begin{center}
	\caption{Sets of parameters in which BHs merged at the masses of the GW events.}
\label{gw_para}
\begin{tabular}{c|c|c|c|c|c|c|c|c|c}
\hline
\multicolumn{10}{c}{{\sf GW150914}} \\
\hline
Model&
$m_0$ ($\mathrm{M_\odot}$)&
$n_\mathrm{gas}$ ($\mathrm{cm^{-3}}$)&
$\epsilon$&
$r_\mathrm{typ}$ (pc)&
$R_\mathrm{gas}$ (pc)&
$m_1$ ($\mathrm{M_\odot}$)&
$m_2$ ($\mathrm{M_\odot}$)&
$t_\mathrm{merge}$ (yr)&
type
\\
\hline
1&25.0&$10^3$&0.1&0.1&10
&37.3&28.2&$3.3\times 10^9$&C\\
\hline
2&25.0&$10^6$&0.01&0.1&1
&33.3&31.3&$2.8\times 10^7$&C\\
\hline
3&25.0&$10^7$&0.01&0.01&0.46
&35.8&32.3&$1.3\times 10^7$&C\\
\hline
4&25.0&$10^8$&0.01&0.01&0.22
&33.8&28.7&$4.2\times 10^5$&C\\
\hline
5&25.0&$10^{10}$&0.01&0.01&0.046
&34.5&32.1&$5.8\times 10^3$&C\\
\hline
6&30.0&$10^4$&0.01&1&4.6
&33.3&31.2&$1.0\times 10^9$&C\\
\hline
7&30.0&$10^4$&0.01&0.1&4.6
&35.6&30.8&$1.0\times 10^9$&C\\
\hline
8&30.0&$10^4$&0.01&0.01&4.6
&33.4&32.0&$7.7\times 10^8$&C\\
\hline
9&30.0&$10^5$&0.01&0.1&2.2
&32.4&31.2&$1.5\times 10^8$&C\\
\hline
10&30.0&$10^6$&0.01&0.1&1
&34.4&32.7&$9.9\times 10^6$&C\\
\hline
11&30.0&$10^9$&0.01&0.1&0.1
&34.4&32.8&$1.7\times 10^4$&C\\
\hline
12&30.0&$10^{10}$&0.001&0.01&0.046
&32.1&31.9&$1.2\times 10^4$&C\\
\hline
\multicolumn{10}{c}{{\sf GW170104}} \\
\hline
Model&
$m_0$ ($\mathrm{M_\odot}$)&
$n_\mathrm{gas}$ ($\mathrm{cm^{-3}}$)&
$\epsilon$&
$r_\mathrm{typ}$ (pc)&
$R_\mathrm{gas}$ (pc)&
$m_1$ ($\mathrm{M_\odot}$)&
$m_2$ ($\mathrm{M_\odot}$)&
$t_\mathrm{merge}$ (yr)&
type
\\
\hline
13&20.0&$10^3$&0.1&1&10
&25.2&22.4&$4.4\times 10^9$&C\\
\hline
14&20.0&$10^8$&0.01&0.1&0.22
&31.3&23.0&$6.1\times 10^5$&C\\
\hline
\multicolumn{10}{c}{{\sf GW170814}} \\
\hline
Model&
$m_0$ ($\mathrm{M_\odot}$)&
$n_\mathrm{gas}$ ($\mathrm{cm^{-3}}$)&
$\epsilon$&
$r_\mathrm{typ}$ (pc)&
$R_\mathrm{gas}$ (pc)&
$m_1$ ($\mathrm{M_\odot}$)&
$m_2$ ($\mathrm{M_\odot}$)&
$t_\mathrm{merge}$ (yr)&
type
\\
\hline
14&20.0&$10^8$&0.01&0.1&0.22
&31.3&23.0&$6.1\times 10^5$&C\\
\hline
15&25.0&$10^5$&0.01&0.1&2.2
&29.9&27.2&$1.2\times 10^8$&C\\
\hline
16&25.0&$10^9$&0.001&0.1&0.1
&29.0&25.0&$8.9\times 10^4$&A\\
\hline
\end{tabular}
\end{center}
\end{table*}

\begin{figure}
\includegraphics[width=85mm]{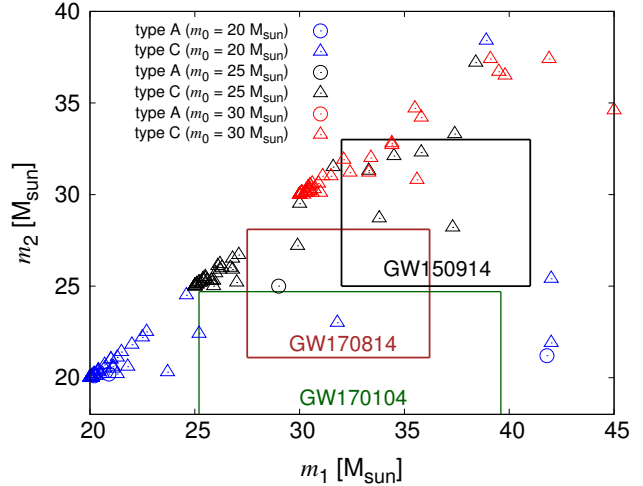}
\caption{BH masses ($m_1>m_2$) 
	in a binary just before the first merger in each run. 
	 Blue, black, and red plots represent 
	 the initial masses ($m_0$) of 20, 25 and 30 $M_\mathrm{\odot}$, respectively. 
	 Circles and triangles represent gas drag-driven mergers (type A) 
	 and three-body driven mergers (type C), respectively. 
	 The masses of GW150914, GW170104, and GW170814 with their uncertainties are 
	 indicated by squares. 
 }
\label{gw_m12}
\end{figure}

\begin{figure}
\includegraphics[width=85mm]{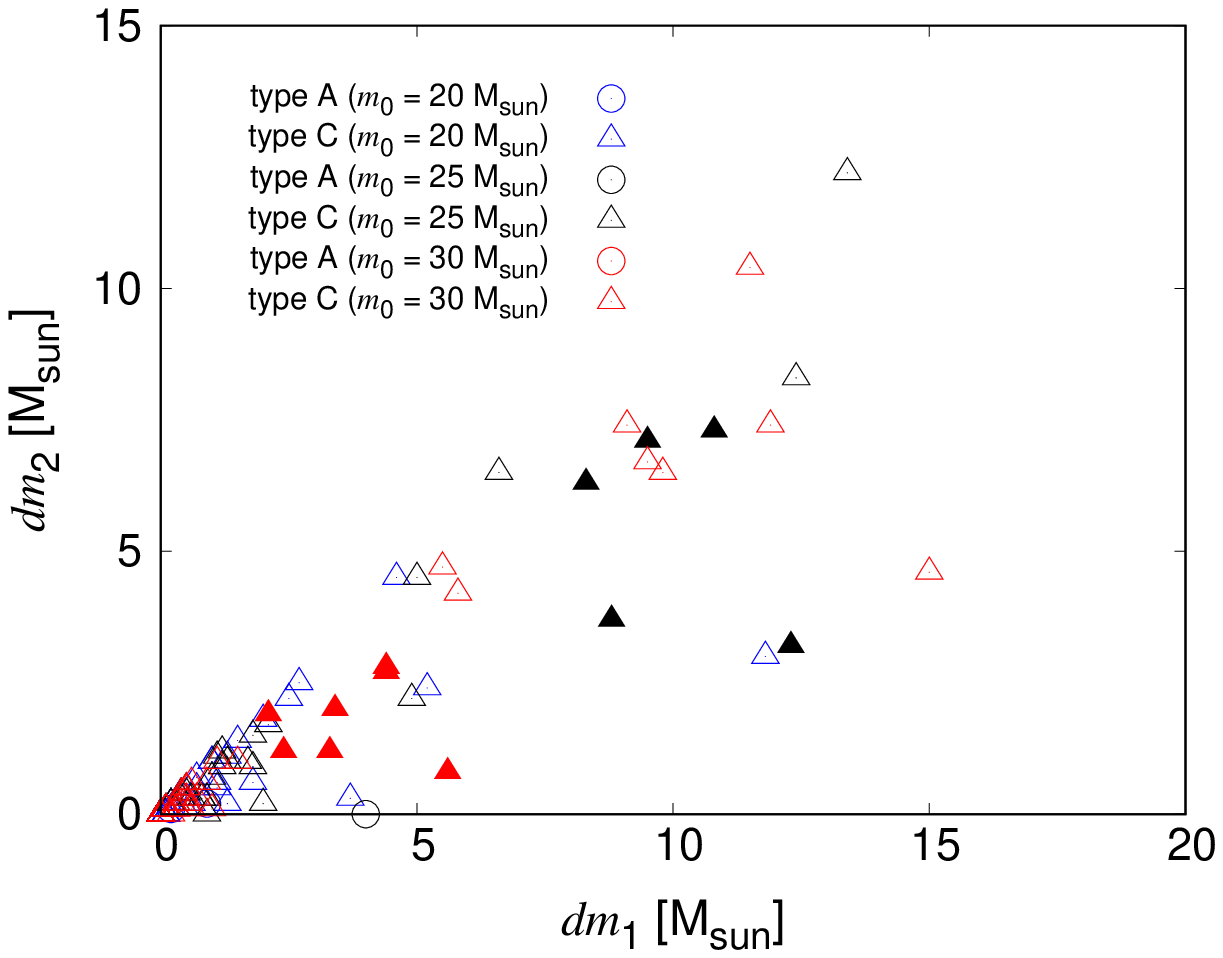}
\caption{Same as Figure \ref{gw_m12}, 
	but accreted masses onto BHs before mergers are shown. 
	 Filled symbols are compatible with the GW150914 event. 
 }
\label{gw_dm12}
\end{figure}

\section{Models Concordant with GW Events}

Changing the set of parameters, we have simulated 264 models, 
of which 135 produce a binary BH merger within 10 Gyr. 
We have found sixteen models to match GW events, where
the final BH masses fall within the estimated mass range in the observations.
In Table \ref{gw_para}, they are listed with the assumed sets of parameters.
The columns are the model number, 
initial mass of BHs ($m_0$), ambient gas number density ($n_\mathrm{gas}$), 
accretion efficiency ($\epsilon$), 
initial extension of BH spatial distributions ($r_\mathrm{typ}$), 
radius of a gaseous sphere ($R_\mathrm{gas}$), 
final masses of merged BHs ($m_1,~m_2,~m_1>m_2$), 
merger time ($t_\mathrm{merge}$), 
and the merger type in each run. 
\citet{tag16} scrutinized merger mechanisms in gas-rich environments. 
They found that gas dynamical friction is indispensable for BH mergers.
First, the BH orbits contract due to gas dynamical friction, 
and then a subsequent merger is promoted through different mechanisms, which are 
classified into four types: 
a gas drag-driven merger (type A), an interplay-driven merger (type B), a three-body driven merger (type C),
and an accretion-driven merger (type D). 

Figure \ref{vtall23} demonstrates the evolution until the first merger in Model 3 
(see Table 1 for the simulation parameters),
where (a) accretion rate, (b) mass and (c) velocity of a heavier BH
in merged BHs, and also (d) separation of the closest pair within all BHs
are shown as a function of time. 
Panels (c) and (d) demonstrate that the velocity decays and
the separation of BHs shrinks owing to gas dynamical friction  
within 2 Myr. 
In this stage, the BH velocity oscillates between subsonic and supersonic one
(the sound speed being $c_{\rm s} = 3.709~{\rm km~s}^{-1}$),
and the accretion rate intermittently reaches a super-Eddington accretion rate. 
In this phase, a binary forms due to energy loss by gas dynamical friction. 
The binary is hardened by kicking another BH through three-body interaction 
at around 2 Myr, which is represented by discontinuous change of the separation. 
Then, the BH velocity becomes highly supersonic and therefore the accretion rate 
is reduced to a level much lower than an Eddington accretion rate. 
A component in the BH binary is sometimes replaced by another one 
as a result of three-body interaction. 
Actually, such exchange occurs at 3.7 and 5.1 Myr.
Three-body interaction is repeated until 13Myr, and 
eventually the binary merges into a massive BH due to GW radiation.
Prior to the BH merger, the masses of merged BHs are enhanced 
by about ten $M_\odot$. As shown in panel (b), 
most of gas accretes in an early three-body interaction phase with subsonic velocity.
The mass accretion rate shortly before the merger emitting GW is 
reduced to less than $10^{-5}$ Eddington accretion rate, owing to the
high circular velocity of the BH binary. 
In practice, the final accretion rate is dependent on ambient gas density.
In the other models listed in Table \ref{gw_para}, the final accretion rate is
$\lesssim 10^{-4}$ Eddington accretion rate. 

In Figure \ref{gw_m12}, we plot the masses of two BHs shortly before first mergers 
in the 135 models out of simulated 264 models, and compare them to
the estimated mass range in the GW150914, GW170104, and GW170814 events. 
We find that the masses of two BHs are consonant to the GW150914 event in twelve models, and
to the GW170104 event in two models, and to the GW170814 event in three models.  
Especially, Model 14 matches the GW170104 and GW170814 events, simultaneously. 
It worth noting that their merger types are type C (three-body driven mergers),
except for Model 16 assuming extremely high density gas.  
Also, it has turned out that, in these twelve models, gas of several $M_\odot$ can accrete onto BHs 
in early three-body interaction phases. 

\citet{Abbott16c} have argued that if a strong stellar wind is assumed,
a BH more massive than $25M_\odot$ should originate in a metal-free (PopIII) or ultra-low metal star.
Even for a weak wind model, the progenitors should be of sub-solar metal abundance.
Hence, the present results imply that metal poor stars are preferred as
the progenitors of the GW150914 BHs. 
Figure \ref{gw_dm12} shows the accumulated mass on each BH before the merger. 
Since three-body interaction is a chaotic process, 
accreting mass in type C changes in a cataclysmic fashion. 
Supposing Bondi accretion, 
there must be uncertainties of $\sim10M_\odot$ in accreting mass, 
since the merger time can fluctuate within a factor of two according to 
the adopted seed random number \citep{tag15}. 
Taking into consideration the fact that the mass uncertainties in the observations are $\sim7$ $M_\odot$, 
about a half of the models that match the masses in the GW150914 event may be missed. 

As shown in Table \ref{gw_para}, the accretion efficiency ($\epsilon$) 
in the concordant models are 0.01, except for the models assuming 
extremely high or low density gas (Model 1, 12, 13 and 16). 
Hoyle-Lyttleton-type accretion is a nonlinear function of mass, 
and therefore the accreting mass is a steep function of $\epsilon$ and $n_\mathrm{gas}$. 
The value of accretion efficiency is roughly determined by the balance 
between accretion timescale and merger timescale \citep{tag16}. 
In other words, the accumulated mass is regulated by these timescales. 
Actually, the timescales accord when the accretion efficiency is around $0.01$. 

\section{Discussion}

\subsection{Merger sites}

We consider preferable sites for the present merger scenario. 
The first possibility is AGN disks, where the density is as high as $\gtrsim10^7~\mathrm{cm}^{-3}$
and the size is as compact as $\lesssim1$ pc \citep{Sirko03, Burtscher13}. 
For a gas disk surrounding a central supermassive BH (SMBH), 
the Toomre $Q$ value is estimated to be
\begin{eqnarray}
Q\simeq 1.4 \left({r \over 1~\mathrm{pc}} \right)^{1/2}
\left( {M_\mathrm{SMBH} \over 10^7~M_\odot} \right)^{1/2}
\left( {M_\mathrm{disk} \over 10^5~M_\odot} \right)^{-1},
\end{eqnarray}
for disk temperature of $10^3~{\rm K}$, where
$M_\mathrm{SMBH}$ and $M_\mathrm{disk}$ is the masses of a SMBH and an AGN disk, respectively. 
Hence, if $M_\mathrm{disk}$ is lower than  $10^5~M_\odot$, the disk is stabilized by the SMBH.
However, a more massive disk should be stabilized by additional heating sources such
as massive stars formed within the disk \citep{Sirko03}.
The viscous timescale of a disk is assessed by
\begin{equation}
t_{\rm vis} \simeq 10^{8}~{\rm yr} 
\left( r \over {\rm 1~pc} \right)^{1/2}
\left( {\alpha \over 0.1} \right)^{-1}
\left( {M_{\rm SMBH} \over 10^{7}~M_\odot} \right)^{1/2},
\end{equation}
where $\alpha$ is the standard viscosity parameter
\citep[e.g.][]{Umemura97}, although the mass accretion 
may be flickering in $\sim 0.1~{\rm Myr}$ \citep{King15}. 
The AGN lifetime can be estimated by the duty cycle, 
$P_{\rm duty}=N_{\rm AGN} t_{\rm AGN} /t_{\rm H}(z)$,
where $t_{\rm AGN}$ is the duration of a single AGN episode,
$N_{\rm AGN}$ is the number of AGN episodes, and
$t_{\rm H}(z)$ is the Hubble time at redshift $z$. 
\citet{Shankar09} have derived $P_{\rm duty}$ as a function of redshift and BH mass.
For $M_{\rm SMBH} =10^7~M_\odot$, 
$P_{\rm duty}\simeq 0.03$ at $z=0.3$ and $P_{\rm duty}\simeq 3\times 10^{-3}$ at $z=0$.
This can be translated into $N_{\rm AGN}t_{\rm AGN}=300$ Myr at $z=0.3$ and 41 Myr at $z=0$, 
while $N_{\rm AGN}t_{\rm AGN}=10$ Myr at $z=0.3$ and 1 Myr at $z=0$
for $M_{\rm SMBH} =10^9~M_\odot$. 

BHs whose orbits are originally misaligned with AGN disks tend to 
be aligned due to gas dynamical friction. 
The alignment timescale is estimated to be
\begin{eqnarray}
t_{\rm align} =
\frac{v_{\rm z}^3}{12\pi G^2 m_0 m_{\rm H}n_\mathrm{gas}} 
\left( {h_{\rm ini} \over h_{\rm disk}}\right)\nonumber\\
\simeq 
10^{8}~{\rm yr} 
\left( {h_{\rm ini} \over 0.2 } \right)^{4}
\left( {h_{\rm disk} \over 0.03 } \right)^{-1}
\left( r \over {\rm 1pc} \right)^{-3/2}\nonumber\\
\left( {n_{\rm gas} \over 10^{7}~{\rm cm}^3} \right)^{-1}
\left( {m_{\rm 0} \over 30~M_\odot} \right)^{-1}
\left( {M_{\rm SMBH} \over 10^{7}~M_\odot} \right)^{3/2},
\label{eq_align}
\end{eqnarray}
where $v_{\rm z}$ is the $z$-component of BH velocity, 
$h_{\rm disk}$ is the aspect ratio of an AGN disk \citep{Goodman03}, 
and $h_{\rm ini}$ is the aspect ratio of an initial BH orbit against a AGN mid-plane. 
Since $t_{\rm align}$ should be shorter than $t_{\rm AGN}$, 
$M_{\rm SMBH}$ is constrained to be $\lesssim 10^{7}~M_\odot$. 
In the process of alignment, the velocity relative to the disk rotation
leads to the epicyclic motion of a BH. The relative velocity is decaying due to dynamical friction,
and simultaneously the circular orbit shrinks in the disk. 
When multiple BHs having residual reciprocal velocity interact with each other in the disk, 
the dynamics similar to the present simulations is expected.  
Also, the situation is analogous to the formation of protoplanets from planetesimals 
in a protoplanetary disk \citep{Kokubo00}. 
\footnote{In practice, the dynamics of multiple BHs in a rotating disk should be explored 
in a more realistic setup, which will be done in the future work.}



Another possibility for the merger site is giant molecular clouds (GMCs). 
The Jeans mass of a cloud with density $n_\mathrm{gas}$ and 
temperature $T$ is 
$M_{\rm J} = 5 \times 10^4~M_\odot (n_\mathrm{gas} /10^3~{\rm cm}^{-3})^{-1/2}(T /10^3~{\rm K} )^{3/2}. $
Therefore, if only thermal pressure is exerted, 
a GMC denser than $10^3~{\rm cm}^{-3}$ is gravitationally unstable
in the free-fall time, $t_{\rm ff}=1.6\times 10^6~{\rm yr}(n_\mathrm{gas} /10^3~{\rm cm}^{-3})^{-1/2}$.
However, GMCs show large non-thermal linewidths indicating supersonic turbulence, 
which may prevent gravitational collapse at large scales \citep[e.g.][]{Boneberg15}.
Actually, the lifetime of GMCs is estimated to be $\sim30$ Myr,
which is longer than the free-fall time \citep[e.g.][]{Krumholz06}. 

Taking into account these timescales in the two possible sites, 
BH mergers should occur within $30-100$ Myr. 
From Table 1, this condition requires $n_\mathrm{gas}\gtrsim10^6~\mathrm{cm}^{-3}$. 
Therefore, dense galactic nuclear disks and dense GMCs 
are potential sites for the mergers concordant with the GW events. 
Besides, dynamically assembled BH binaries in the present simulations
predict an isotropic distribution of spin tilts without alignment with the orbital angular momentum, 
which is preferred to account for the misaligned spins in the GW170104 event.

\subsection{Event rate in AGNs}

We estimate the event rates for mergers of massive stellar-mass BHs
in the first advanced LIGO observation run (LIGO O1). 
The horizon distance of massive BH mergers is 
$D_h\approx3~\mathrm{Gpc}~(z\approx0.3)$ \citep[]{Belczynski16}, 
corresponding to a comoving volume $V_c \approx 50~\mathrm{Gpc}^3$. 
Here, we assess the event rates for massive BH mergers in AGN gas disks. 

First, we consider remnant BHs of massive population I stars 
formed in a galaxy, say, PopI BHs. 
Due to inward migration of BHs by stellar dynamical friction, 
about $N_\mathrm{BH}\sim2\times10^4$ BHs may exist within $1$ pc from a SMBH 
in a Milky Way (MW)-sized Galaxy \citep*{Miralda00,Antonini14}. 
To produce massive BHs with $\gtrsim25~M_\odot$, the initial progenitor mass is required to be 
$\gtrsim70~M_\odot$ \citep{Belczynski16}. 
Supposing the Salpeter initial mass function with an upper mass limit of
$100M_\odot$, $\sim 20\%$ of produced BHs are expected to be massive
($f_\mathrm{massive}\sim 0.2$). Hence, the fraction of massive BH pairs is
$f_\mathrm{massive}^2\sim 0.04$.
Since the aspect ratio ($h_{\rm ini}$) represents the fraction of BHs which can align to the AGN disk,  
the fraction of aligned BHs in $t_{\rm AGN}$ is given by 
$f_{\rm align}\simeq 0.2(t_{\rm AGN}/100~{\rm Myr})^{1/4}$ from equation (\ref{eq_align}).
In order for a merger to take place, the condition of 
$t_{\rm merge} \le t_{\rm AGN}$ should be satisfied, where 
is $t_{\rm merge} \sim 10~{\rm Myr}$ for $n \sim 10^7~{\rm cm}^{-3}$ from the present simulations.
Thus, it is required that $t_{\rm AGN} \ge 10~{\rm Myr}$ 
and therefore $N_{\rm AGN} \le P_{\rm duty}t_{\rm H}(z)/10~{\rm Myr}$.
In the range of $M_{\rm SMBH} \leq 10^7~M_\odot$ and $0 \lesssim z \lesssim 0.3$,
we have $3\times 10^{-3} \lesssim P_{\rm duty} \lesssim 3\times 10^{-2} $ \citep{Shankar09}.
Then, $N_{\rm AGN} \le 4$ at $z \sim 0$ and $N_{\rm AGN} \le 31$ at $z \sim 0.3$.
Using these assessments, the merger rate per Milky-sized galaxy is estimated to be 
$\dot{N}_\mathrm{merge/gal} \sim 
P_{\rm duty} f_{\rm align} f_{\rm massive}^2 N_{\rm BH} /t_{\rm AGN} 
=f_{\rm align} f_{\rm massive}^2 N_{\rm BH} N_{\rm AGN}/t_{\rm H}(z) 
\simeq 10 - 20~{\rm Gyr}^{-1}$ for  $N_{\rm AGN}=1$, and 
$\simeq 30 - 300~{\rm Gyr}^{-1}$ for the maximum of $N_{\rm AGN}$.
From the Schechter function fit of local galaxies, 
the number density of MW-sized galaxies is $n_\mathrm{gal}\sim 2\times 10^6~\mathrm{Gpc}^{-3}$ \citep{Marzke98}. 
Using these values, the number of MW-sized galaxies involved in an observable volume 
is $N_\mathrm{gal}\sim V_c n_\mathrm{gal}\sim1\times10^8$. 
Under these assumptions, the event rate for mergers of massive PopI BHs 
in AGN disks in the first observing run of aLIGO is estimated to be 
$R_\mathrm{O1,AGN,PopI}\sim \dot{N}_\mathrm{merge/gal}
N_\mathrm{gal}\simeq 1-2~\mathrm{yr}^{-1}$ for  $N_{\rm AGN}=1$,
and
$\simeq 3 - 30~{\rm yr}^{-1}$ for the maximum of $N_{\rm AGN}$.
The volumetric event rate is 
$R_\mathrm{vol,AGN,PopI}\sim \dot{N}_\mathrm{merge/gal} n_\mathrm{gal}
\simeq (2-4) \times 10^{-2}~\mathrm{Gpc}^{-3}\mathrm{yr}^{-1}$ for  $N_{\rm AGN}=1$,
and
$ \simeq 0.06-0.6~\mathrm{Gpc}^{-3}\mathrm{yr}^{-1}$ for the maximum of $N_{\rm AGN}$.

Next, we consider remnant BHs of population III stars (PopIII BHs). 
Although there are large uncertainties, 
roughly ten PopIII BHs are possibly born in a minihalo of $10^5-10^6~M_\odot$ \citep{sus14,Valiante16}. 
In this case, $\sim10^6$ PopIII BHs are expected to exist in a MW-sized galaxy \citep{Ishiyama16}. 
Then, if the ratio of PopIII BHs to PopI BHs is assumed to be constant in a whole galaxy, 
the number of BHs in central subparsec regions is $N_\mathrm{BH}\sim 2\times 10^2$. 
Besides, if taking into consideration the possibility that BHs within $\sim 10$ pc can migrate into subparsec regions, 
the number of BHs at $\lesssim 1$ pc can increase by about one order of magnitude \citep{Miralda00}. 
So, we suppose $N_\mathrm{BH}\sim 2 \times10^{3}$ PopIII BHs exist in an AGN disk in a MW-sized galaxy. 
We assess the fraction of massive ones in all PopIII BHs to be $f_\mathrm{massive}\sim0.5$ \citep*{Heger02,sus14}.
Then, 
$\dot{N}_\mathrm{merge/gal} \sim 
P_{\rm duty} f_{\rm align} f_{\rm massive}^2 N_{\rm BH} /t_{\rm AGN} 
\simeq 6 - 10~{\rm Gyr}^{-1}$ for  $N_{\rm AGN}=1$, and 
$\simeq 20 - 200~{\rm Gyr}^{-1}$ for the maximum of $N_{\rm AGN}$.
Under these assumptions, 
we estimate the event rate for mergers of massive PopIII BHs in AGN disks to be 
$R_\mathrm{O1,AGN,PopI}\sim \dot{N}_\mathrm{merge/gal}
N_\mathrm{gal}\simeq 1~\mathrm{yr}^{-1}$ for  $N_{\rm AGN}=1$,
and
$\simeq 2-20~{\rm yr}^{-1}$ for the maximum of $N_{\rm AGN}$.
The volumetric event rate is 
$R_\mathrm{vol,AGN,PopI}\sim \dot{N}_\mathrm{merge/gal} n_\mathrm{gal}
\simeq 2 \times 10^{-2}~\mathrm{Gpc}^{-3}\mathrm{yr}^{-1}$ for  $N_{\rm AGN}=1$,
and 
$\simeq 0.04-0.4~\mathrm{Gpc}^{-3}\mathrm{yr}^{-1}$ for the maximum of $N_{\rm AGN}$.  

\subsection{Event rate in GMCs}

Here, we estimate the event rates for BH mergers in GMCs. 
A MW-sized galaxy contains $\sim10^8$ BHs in the volume of
$\sim100~\mathrm{kpc}^3$ \citep{rem06}, where the fraction of massive BHs
is 0.2 as discussed in previous section.
There are $\sim1000$ GMCs in a galaxy and
they occupy the volume $10^{-3}~\mathrm{kpc}^3$ \citep{Ruffle06}. 
Hence, we can assume that $\sim 200$ massive BHs reside in GMCs. 
Considering the fact that 
velocity dispersion of PopI massive stars is $\sim20~\mathrm{km/s}$ \citep*{Binney98,Nordstrom04}
and the escape velocity of GMCs of $\sim10~\mathrm{km/s}$ \citep*{Dobbs11,Dale12}, 
$\sim40$ percent of PopI BHs can be captured by GMCs. 
According to probability distributions, 
about 3 GMCs possess more than two massive BHs.  
Also, the volume filling factor of dense cores in GMCs is $f_{\rm core} \sim 0.05$ \citep{Bergin96}. 
Besides, stars which leave BHs more massive than $25~M_\odot$ 
should be metal poor ($\leq 0.3$ solar metallicity)
and they should be low velocity dispersion \citep{Nordstrom04}. 
Most of such stars exist in outer galaxies of $\gtrsim 10$ kpc \citep{Martinez-Medina17}, 
where the stellar mass is $\sim 0.1$ of the total galactic stellar mass. 
Since BHs are redistributed in the dynamical time of a galaxy $t_\mathrm{dyn}\sim 100$ Myr, 
the merger rate in a MW-sized galaxy is 
$\dot{N}_\mathrm{merge/gal}\sim 3\times 0.1 f_{\rm core}/t_\mathrm{dyn}\simeq 0.2 ~\mathrm{Gyr}^{-1}$. 
Under these assumptions, the event rate for mergers of PopI BHs in GMCs 
is estimated to be 
$R_\mathrm{O1,GMC,PopI} \sim \dot{N}_\mathrm{merge/gal} N_\mathrm{gal}\simeq 0.02~\mathrm{yr}^{-1}$ 
and 
$R_\mathrm{vol,GMC,PopI} \sim \dot{N}_\mathrm{merge/gal} n_\mathrm{gal}
\simeq 3\times 10^{-4}~\mathrm{Gpc}^{-3}\mathrm{yr}^{-1}$. 



\section{Conclusions}

In this paper, we have considered the mergers of unassociated BHs
through a chance meeting in gas-rich environments.
To elucidate the merger condition 
concordant with the recently detected gravitational wave events, 
we have conducted highly accurate post-Newtonian $N$-body simulations
on a multiple BHs system embedded in dense gas, 
incorporating dynamical friction, Hoyle-Lyttleton mass accretion, 
and general relativistic effects such as pericentre shift and gravitational wave emission. 
Consequently, we have found the following: 

\begin{enumerate}
		\renewcommand{\labelenumi}{(\arabic{enumi})}

\item \noindent
Gas dynamical friction works effectively to promote three-body interaction of BHs
in dense gas of $n_\mathrm{gas}\gtrsim 10^6 ~\mathrm{cm}^{-3}$. 
Eventually, mergers are caused within  $30$ Myr. 
This scenario predicts an isotropic distribution of spin tilts, which is compatible with
the spin misalignment seen in the GW170104 event. 

\item \noindent
Before BH mergers, gas of several $M_\odot$ accretes onto each BH.
However, gas accretion takes place predominantly during early three-body interaction phases,
and the final mass accretion rates shortly before GW emmision are
$\lesssim10^{-4}$ Eddington accretion rate.
Thus, the electromagnetic counterparts of GW events might not be so luminous.

\item \noindent
We have found sets of model parameters concordant 
with the massive BHs detected in the GW events. 
In the concordant models, the initial extension of BH distributions is smaller than $1$ pc.
To account for the GW150914 event,
the masses of seed BHs are required to be $\gtrsim25M_\odot$.
Hence, metal poor stars are preferred as the progenitors of the GW150914 BHs. 

\item \noindent
We have roughly estimated the event rates by the first observing run of LIGO advanced detectors. 
The event rates for massive PopI BHs and PopIII BHs in AGN disks are assessed to be
$\simeq 1-2~\mathrm{yr}^{-1}$ and $\simeq 1~\mathrm{yr}^{-1}$, respectively.
If multiple episodes of AGNs are taken into consideration, the rates can be enhanced by 
roughly an order of magnitude. 
For massive PopI BHs in dense interstellar clouds, the rate is $\simeq 0.02~\mathrm{yr}^{-1}$. 
Hence, high-density AGN disks are a more plausible site for
mergers of chance meeting BHs. 

\end{enumerate}

In the present simulations, we have assumed a fairy simple configuration of matter. 
However, taking realistic situations into consideration, 
we should construct a more concrete model of gas distributions in a dense cloud/disk
and gravitational potential, including stellar distributions 
and a central supermassive black hole. 
Also, the back reaction due to gas dynamical friction 
may alter the BH dynamics. These effects will be explored
in the future analysis. 


\acknowledgments

We thank the anonymous referee for useful comments. 
Numerical computations and analyses were carried out on Cray XC30 
and computers at Center for Computational Astrophysics, 
National Astronomical Observatory of Japan, respectively. 
This research is also supported in part by 
the European Research Council under the European Unionfs Horizon 2020 Programme, 
ERC-2014-STG grant GalNUC 638435 and 
Interdisciplinary Computational Science Program 
in Center for Computational Sciences, University of Tsukuba, 
and Grant-in-Aid for Scientific Research (B) by JSPS (15H03638).


\clearpage



\end{document}